\documentclass[lettersize,journal]{IEEEtran}
\usepackage[utf8]{inputenc}
\usepackage{amsmath,amsfonts}
\usepackage{xcolor}
\usepackage{algorithm}
\makeatletter
\renewcommand\fs@ruled{\def\@fs@cfont{\bfseries}\let\@fs@capt\floatc@ruled
  \def\@fs@pre{\normalcolor\hrule height.8pt depth0pt \kern2pt}%
  \def\@fs@post{\kern2pt\normalcolor\hrule\relax}%
  \def\@fs@mid{\kern2pt\normalcolor\hrule\kern2pt}%
  \let\@fs@iftopcapt\iftrue} %
\makeatother
\floatstyle{ruled}
\restylefloat{algorithm}
\usepackage{array}
\usepackage{textcomp}
\usepackage{stfloats}
\usepackage{url}
\usepackage{verbatim}
\usepackage{graphicx}
\usepackage{stfloats}
\usepackage{cite}
\usepackage{algpseudocode}
\usepackage[font=small,labelsep=period]{caption}
\usepackage[subrefformat=parens]{subcaption}
\algrenewcommand\algorithmicrepeat{\textbf{Repeat}}
\algrenewcommand\algorithmicuntil{\textbf{Until}}
\begin{document}

\title{Distributed Electromagnetic Neural Networks for Task-Oriented Semantic Communications}

\author{Jinbao Li, Jiancheng An,~\IEEEmembership{Senior Member,~IEEE,} \\ Hao Liu,~\IEEEmembership{Graduate Student Member,~IEEE,} Lu Gan,~\IEEEmembership{Member,~IEEE,} \\Victor C. M. Leung,~\IEEEmembership{Life Fellow,~IEEE,} Mehdi Bennis,~\IEEEmembership{Fellow,~IEEE,} and M\'{e}rouane Debbah,~\IEEEmembership{Fellow,~IEEE}

\thanks{This work is supported by National Natural Science Foundation of China 62471096. (\emph{Corresponding Author: Jiancheng An})}%
\thanks{J. Li, H. Liu, and L. Gan are with the School of Information and Communication Engineering, University of Electronic Science and Technology of China (UESTC), Chengdu, 611731, China (e-mail: li.jinbao@std.uestc.edu.cn; liu.hao@std.uestc.edu.cn; ganlu@uestc.edu.cn).}
\thanks{J. An is with the School of Electronic Science and Engineering, University of Electronic Science and Technology of China (UESTC), Chengdu, 611731, China (e-mail: jiancheng\_an@163.com).}
\thanks{V. C. M. Leung is with the Department of Electrical and Computer Engineering, The University of British Columbia, Vancouver, BC V6T 1Z4, Canada (e-mail: vleung@ieee.org).}
\thanks{M. Bennis is with the Centre for Wireless Communications, University of Oulu, Oulu 90014, Finland (e-mail: mehdi.bennis@oulu.fi).}
\thanks{M. Debbah is with the Research Institute for Digital Future, Khalifa University, 127788 Abu Dhabi, UAE (email: merouane.debbah@ku.ac.ae).}\vspace{-0.8cm}
}
\IEEEpubid{}
\maketitle

\begin{abstract}
Semantic communications (SemCom) is a promising paradigm that prioritizes the transmission of task-relevant information, thereby enabling superior communication efficiency over traditional bit-centric systems. However, most existing SemCom systems face critical limitations in computational efficiency and spatial flexibility. To overcome these limitations, we propose a novel unmanned aerial vehicles (UAV)-enabled distributed electromagnetic neural network (EMNN) for a task-oriented SemCom system. Specifically, the proposed distributed EMNN is composed of multiple UAV-mounted stacked intelligent metasurfaces (SIM) and a ground receiving station (GRS), where multiple SIMs collaboratively encode image semantics in the wave domain, and the GRS performs decoding based on the received power distribution. Moreover, we employ a temperature-adaptive gradient optimization algorithm to train the distributed EMNN, which mitigates gradient vanishing and enhances learning stability. Finally, the numerical simulation results demonstrate the effectiveness of distributed EMNN in image recognition task-oriented SemCom, achieving an average $8\%$ accuracy improvement over the single-SIM baseline across multiple datasets.
\end{abstract}


\vspace{-0.4cm}
\section{Introduction}
\IEEEPARstart{S}{emantic} communications (SemCom) have emerged as a promising approach to meet the demands of next-generation networks, which require low latency and reduced bandwidth consumption. By transmitting task-relevant information rather than raw bits, SemCom has demonstrated superior efficiency over traditional methods in tasks such as image recognition and natural language understanding \cite{bennis2025semantic}. Despite recent advances, most existing SemCom systems are fundamentally dependent on digital neural networks to encode and decode semantic information, which incur substantial computational resources, power consumption, and processing latency \cite{Liu2025WCM, Huang2024SemComTVT}, thus hindering both real-time deployment and scalability, especially in power-constrained environments like unmanned aerial vehicles (UAVs) or mobile edge devices.
 
To tackle these challenges, stacked intelligent metasurfaces (SIM) have emerged, which offloads semantic processing from digital to the wave domain. Specifically, an SIM consists of multiple closely parallel metasurface layers, where each layer is composed of a dense array of meta-atoms. The transmission coefficients of meta-atoms can be dynamically adjusted by field-programmable gate arrays (FPGA) in real time, enabling programmable wavefront manipulation \cite{an2024stacked}. The multi-layer structure allows the SIM to function as an electromagnetic neural network (EMNN), where the meta-atoms act as electromagnetic (EM) neurons, collectively performing direct transformations on input EM waves \cite{lin2018all}. In contrast to conventional digital processing, SIM enables high-speed inference in the analog wave domain, leading to significantly lower energy consumption and processing latency \cite{liu2022programmable}.

Recent studies have demonstrated the superiority of SIM for executing diverse computational functions across wireless communication and sensing domains \cite{Nad2024hyb, lin2025uav, Shi2025uplink}. Building on these, subsequent advances have extended these architectures into EMNNs. Specifically, Huang et al. employed a centralized EMNN for SemCom to perform wave-domain semantic encoding \cite{huang2024stacked}. However, this single-point deployment lacks spatial flexibility and remains highly vulnerable to dynamic environments. Meanwhile, distributed intelligent metasurfaces \cite{DistributedRIS} have been investigated to enhance propagation reliability. Nevertheless, they typically function as single-layer relays focusing on channel reconfiguration, lacking the architectural depth required for semantic feature extraction. In contrast, a distributed EMNN deployment with multiple UAVs leverages spatial diversity for robust semantic feature extraction and utilizes over-the-air computation (AirComp) to naturally aggregate these distributed waveforms through the wireless channel. Such a flexible framework ensures critical resilience in challenging environments, such as disaster zones or congested urban centers \cite{alotaibi2025disaster}. Even under severe localized interference, the remaining spatially separated nodes can collaboratively maintain reliable semantic delivery via the over-the-air wave-domain fusion.

Based on the above observations, we propose a distributed EMNN for a task-oriented SemCom system. The system comprises multiple UAV-mounted SIMs acting as source and semantic encoders and an antenna array at the ground receiving station (GRS) functioning as a decoder based on the received power distribution. To achieve the desired function, we propose a temperature-adaptive gradient optimization (TAGO) algorithm to train the distributed EMNN, which introduces a learnable temperature parameter to stabilize the training process and improve performance. Finally, numerical results validate the effectiveness of the proposed EMNN and algorithm in the task-oriented SemCom system. 

\begin{figure}[t!]
 \centering
 \includegraphics[width=0.45\textwidth]{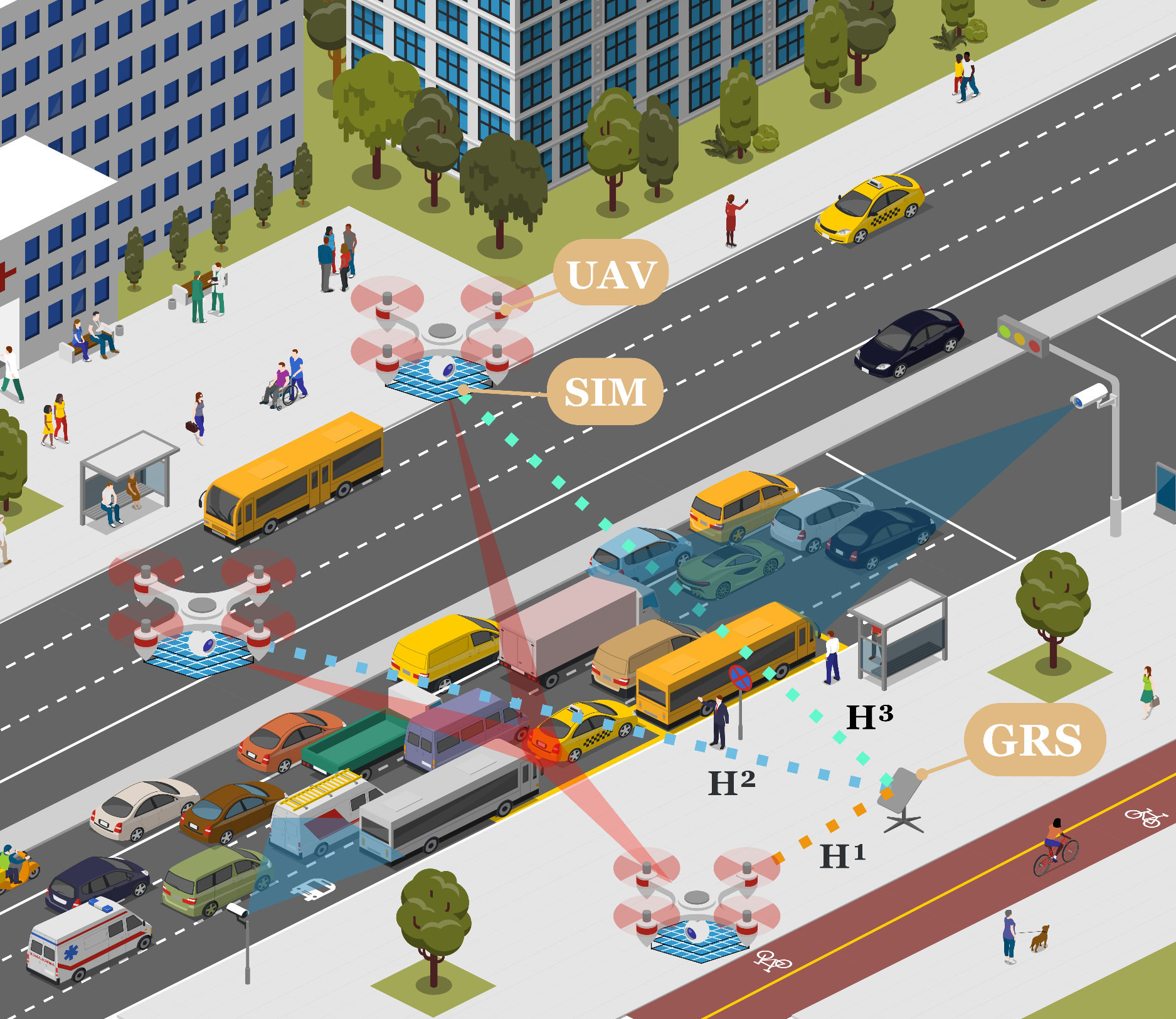} 
 \caption{Task-oriented SemCom system enabled by UAV-mounted distributed EMNN, exemplified by collaborative license plate detection under severe occlusions in congested urban traffic.}
 \label{fig:system_model}
 \vspace{-0.6cm}
\end{figure}

\vspace{-0.4cm}
\section{System Model}
In this section, we present a task-oriented SemCom system enhanced by distributed EMNN. Specifically, Section \ref{subsec:uav_tx} introduces the UAV-side source and semantic encoder, Section \ref{subsec:channel} describes the wireless channel model, and Section \ref{subsec:grs_rx} outlines the semantic decoding mechanism at the GRS.

\vspace{-0.4cm}
\subsection{UAV-Side Source and Semantic Encoder} \label{subsec:uav_tx}
In this letter, we consider an image recognition task-oriented semantic system. As shown in Fig. \ref{fig:system_model}, multiple UAV-mounted SIMs are deployed in a distributed manner to collaboratively perform source and semantic encoding in the wave domain, ensuring that task-relevant semantic features are jointly extracted and delivered to the GRS. On the UAV side, as the primary component of EMNN, the SIM consists of $L$ layers, each implemented as a uniform planar array (UPA) composed of \(N=N^{\text{row}}N^{\text{col}}\) independently tunable meta-atoms, where \(N^{\text{row}}\) and \(N^{\text{col}}\) denote the number of meta-atoms along rows and columns. The indices of different UAV-mounted SIMs are represented by \(\mathcal{K}=\left \{1,2,\dots,K\right \} \), the layers within each SIM are indexed by \(\mathcal{L}=\left \{ 1,2,\dots ,L \right \} \), and the meta-atoms within each layer are indexed by \(\mathcal{N}=\left \{ 1,2,\dots ,N \right \}\). Meanwhile, the receiving antennas at the GRS are indexed by \(\mathcal{M}=\left \{ 1,2,\dots ,M \right \}\). The incident EM waves from the antenna undergo amplitude and phase modulation at each meta-atom when passing through a metasurface, and the resulting modulated waves serve as secondary sources that illuminate meta-atoms on the subsequent layer. Thus, SIM emulates a fully connected EMNN in the wave domain. We define the trainable coefficient of the $n$-th meta-atom on the \(l\)-th layer of the $k$-th SIM as \(\psi_{l,n}^{k} = \alpha _{l,n}^{k} e^{j\theta _{l,n}^{k} } \), \( k \in \mathcal{K}\), \( l \in \mathcal{L}\setminus \left \{ 1 \right \}\), \( n \in \mathcal{N}\), where \(\alpha _{l,n}^{k} \in \left [ 0,1 \right ] \) and \(\theta _{l,n}^{k} \in \left [ 0,2\pi \right )\) denote its amplitude and phase shift, respectively. The corresponding transmission coefficient vector and matrix can be defined as $\boldsymbol{\psi}^{k}_{l}=\ [ \psi^{k}_{l,1}, \psi^{k}_{l,2}, \dots, \psi^{k}_{l,n} ]^T \in \mathbb{C}^{N\times 1}$ and \(\mathbf{\Psi}^{k}_{l}=\text{diag}\left ( \boldsymbol{\psi}^{k}_{l}\right ) \in \mathbb{C}^{N\times N}\), respectively. 

Moreover, as all SIMs are assumed to possess an identical structural configuration, the inter-layer transmission matrix across different SIMs can be uniformly denoted as $\textbf{W}_{l} \in \mathbb{C}^{N \times N}$, \( l \in \mathcal{L}\setminus \left \{ 1 \right \}\), characterizing wave propagation between neighboring layers of the SIM. Its $(n,n^\ast)$-th entry, $w_{l, (n,n^\ast)}$, specifies the transmission coefficient from the $n^\ast$-th meta-atom on the $(l-1)$-th layer to the $n$-th meta-atom on the $l$-th layer, which can be defined by Rayleigh-Sommerfeld diffraction theory as \cite{liu2022programmable}
\vspace{-0.3cm}
\begin{equation}
\label{transmission matrix}
w_{l, (n, {n}^{\ast})}=\frac{d_{L}S}{\left(d_{l, (n, {n}^{\ast})}\right)^{2}}\left(\frac{1}{2 \pi }+ \frac{d_{l, (n, {n}^{\ast})}}{j\lambda }\right) e^{\frac{j 2 \pi d_{l, (n, {n}^{\ast})}}{\lambda}},
\end{equation}
where \(\lambda\) and \(S\) are the wavelength of the carrier signal and the area of each meta-atom, respectively. \(d_{L}=T_{\text{SIM}}/(L-1)\) represents the spacing between two adjacent layers, with \(T_{\text{SIM}}\) denoting the thickness of the SIM. \(d_{l, (n, {n}^{\ast})}\) denotes the propagation distance between the considered two meta-atoms. 

Within SIM, the encoding process is divided into two stages: the source encoding layer, mapping the original image information into the EM signal, and the semantic encoding layers, refining the semantic features through multi-layer modulation. 
\subsubsection{Source Encoding Layer}
The first metasurface of the SIM functions as an image encoding interface, mapping image pixels into meta-atoms' amplitude-phase profiles. For grayscale images with one channel, we assume \( \alpha _{1,n}^{k}=1 \) and map the grayscale channel to the phase component $\theta_{1,n}^k$. When extending to colored images, we adopt the luminance (Y) channel and the combination of subsampled chrominance (Cb and Cr) channels, respectively, configure the real and imaginary components, i.e., $\left | \alpha _{1,n}^{k}\right |\cos\theta_{1,n}^{k}$ and $\left | \alpha _{1,n}^{k}\right |\sin\theta_{1,n}^{k}$, of the transmission coefficients. Subsequently, the transmitting antennas of UAVs emit EM waves that carry the image information $\mathbf{\Psi}^k_1$ from the first layer to the subsequent layers for semantic encoding. As the transmitting antenna is located close to the source encoding layer, the corresponding propagation vector $\textbf{w}_{1} \in \mathbb{C}^{N \times 1}$ can also be characterized by (\ref{transmission matrix}).
\subsubsection{Semantic Encoding Layers}
After completing source encoding, the subsequent metasurfaces act as a semantic encoder. According to the Huygens–Fresnel principle \cite{lin2018all}, the semantic encoding process in the $k$-th SIM can be expressed as
\begin{equation}
\label{eq:G}
\textbf{G}^k=\mathbf{\Psi}_{L} ^k\textbf{W}_{L}\mathbf{\Psi}_{L-1}^{k}\textbf{W}_{L-1}\cdots\mathbf{\Psi}_{2}^{k}\textbf{W}_{2}\in\mathbb{C}^{N\times N}.
\end{equation}
Through successive multi-layer modulation in (\ref{eq:G}), SIM progressively achieves semantic encoding. Thus, by leveraging $K$ UAV-mounted SIMs at spatially diverse locations, a distributed EMNN is formed to collaboratively extract task-relevant features, improving inference robustness and accuracy.

\vspace{-0.5cm}
\subsection{Wireless Channel} \label{subsec:channel}
The channel from output metasurface of the $k$-th SIM to the GRS can be represented by \(\mathbf{H}^k=\left[\mathbf{h}^k_1, \mathbf{h}^k_2, \ldots, \mathbf{h}^k_M \right]^{T} \in \mathbb{C}^{M \times N}\), where \(\textbf{h}^k_m \in \mathbb{C}^{N\times1}\), \(k \in \mathcal{K}, m \in \mathcal{M}\) represents the channel from the $m$-th antenna of the GRS to the output layer of the $k$-th SIM. Specifically, we adopt a Rician fading channel \cite{liu2025Novel} characterized by
\begin{equation}
\label{wireless_channel}
\mathbf{H}^k=\sqrt{\frac{\Omega^{k} }{1+Z_{H}^{k}}} (\sqrt{Z_{H}^{k}}\textbf{H}^{k}_{\text{LoS}}+\textbf{H}^{k}_{\text{NLoS}}),
\end{equation}
where \(\Omega^{k} \) and \(Z_{H}^{k}\) represent the path loss and Rician factor, associated with the $k$-th SIM. \(\textbf{H}^{k}_{\text{LoS}}\) is the line-of-sight (LoS) component, while \(\textbf{H}^{k}_{\text{NLoS}}\) denotes the non-LoS (NLoS) component of the $k$-th channel. In this letter, we assume that the UAVs maintain stable hovering positions during each short semantic transmission frame. Under this quasi-static condition, the Rician factor and the corresponding LoS and NLoS components remain constant within a single transmission block.

At the GRS, we assume that the UPA antenna array comprises \(M=M_{X}M_{Y}\) antenna elements, uniformly spaced by a distance \(d_{M}\) along both the $x$-axis and the $y$-axis. Let \(\mu^k \in \left [ 0,2\pi \right ) \) and \(\beta^k \in \left [ 0,\pi/2 \right ]\) denote the physical azimuth and elevation angles of arrival associated with the $k$-th UAV, respectively. Thus, the electrical angles \(\eta ^{k,x}_R \) and \(\eta ^{k,y}_R \) along the $x$-axis and $y$-axis are respectively denoted by
\vspace{-0.1cm}
\begin{equation}
\label{eta_x}
\eta ^{k,x}_R=\frac{2\pi d_{M} }{\lambda } \sin (\beta^k)\cos (\mu^k),
\end{equation}
\vspace{-0.3cm}
\begin{equation}
\label{eta_y}
\eta ^{k,y}_R=\frac{2\pi d_{M}}{\lambda } \sin (\beta^k)\sin (\mu^k).
\end{equation}
Let \(\textbf{v}^{k,x}=\left [ 1,e^{j\eta ^{k,x}_R},\dots,e^{j\eta ^{k,x}_RM_{X}} \right ]^T \in \mathbb{C}^{M_{X}\times1} \) and \(\textbf{v}^{k,y}=\left [ 1,e^{j\eta ^{k,y}_R},\dots,e^{j\eta ^{k,y}_RM_{Y}} \right ]^T \in \mathbb{C}^{M_{Y}\times1} \). Thus, the steering vector \(\textbf{b}_{R}^k(\eta ^{k,x}_R,\eta ^{k,y}_R)\) of the GRS can be defined as
\begin{equation}
\label{a_g}
\textbf{b}_{R}^k(\eta ^{k,x}_R,\eta ^{k,y}_R)= \textbf{v}^{k,x}\otimes \textbf{v}^{k,y} \in \mathbb{C}^{M\times1},
\end{equation}
where \(\otimes\) denotes the Kronecker product. Analogously, the steering vector \(\textbf{b}_{S}^k(\eta ^{k,x}_{S},\eta ^{k,y}_{S}) \in \mathbb{C}^{N\times1}\) of the output metasurface of the $k$-th SIM can be obtained in the same manner. Thus, \(\textbf{H}_{\text{LoS}}^k\) can be described as follows:
\begin{equation}
\label{H_LoS}
\textbf{H}_{\text{LoS}}^k=\textbf{b}_{R}^k(\eta ^{k,x}_R,\eta ^{k,y}_R)\textbf{b}_{S}^k(\eta ^{k,x}_{S},\eta ^{k,y}_{S})^H \in \mathbb{C}^{M\times N}.
\end{equation}

The NLoS component follows the circularly symmetric complex Gaussian distribution with zero mean and an identity covariance matrix, i.e., vec\{$\mathbf{H}^{k}_{\text{NLoS}}\} \sim \mathcal{C N}(0,\textbf{I}_{MN})$, where \(\textbf{I}_{MN} \in \mathbb{C}^{MN\times MN} \) denotes the identity matrix.

Hence, the signal received at the GRS through the $k$-th SIM can be modeled as
\vspace{-0.1cm}
\begin{equation}
\label{x}
\textbf{x}^k=\sqrt{p_t}\mathbf{H}^k\mathbf{G}^k\mathbf{\Psi}_{1}^k\mathbf{w}_{1}+\textbf{n}^k\in \mathbb{C}^{M\times 1},
\end{equation} 
where \(p_t\) represents the transmit power at the UAV and $\mathbf{n}^k$ is the additive white Gaussian noise at the receiver, satisfying $\mathbf{n}^k\sim\mathcal{C N}(0,\sigma^{2}\textbf{I}_{M})$, where $\sigma^{2}$ is the average noise power.

\vspace{-0.5cm}
\subsection{GRS-Side Semantic Decoder} \label{subsec:grs_rx}
The decoder at the GRS features a UPA with $M$ antennas, each corresponding to one semantic category. The classification is achieved by aggregating signals from multiple SIMs and determining which antenna receives the maximum signal power. In this letter, we consider two combining strategies at the GRS: pre-detection combining and post-detection combining. The pre-detection strategy fundamentally implements AirComp by orchestrating concurrent transmissions from multiple UAVs, which transforms the wireless channel into an analog fusion layer for natural wave-domain signal superposition. Since spatially distributed UAVs experience independent fading paths, the over-the-air aggregation directly harnesses spatial diversity gain, effectively mitigating localized deep fading. The composite signal can be defined as:
\vspace{-0.1cm}
\begin{equation}
\label{s1}
\hat{\textbf{y}} = \left | \sum_{k=1}^{K} \textbf{x}^{k} \right | ^{2} \in \mathbb{R}^{M\times 1},
\end{equation}
while the signal through post-detection combining can be expressed as
\vspace{-0.5cm}
\begin{equation}
\label{s2}
\hat{\textbf{y}} = \sum_{k=1}^{K}\left | \textbf{x}^{k} \right | ^{2} \in \mathbb{R}^{M\times 1}.
\end{equation}
Notably, the modulo operations in (\ref{s1}) and (\ref{s2}) serve as nonlinear activations that break the linear cascade bottleneck, enabling effective semantic inference for real-world tasks.

\begingroup
\setlength{\textfloatsep}{-0.6cm}
\begin{algorithm}[t] 
\color{black}
\caption{Temperature-Adaptive Gradient Optimization (TAGO) Algorithm for Training Distributed EMNN} 
\begin{algorithmic}[1]
\setlength{\itemsep}{2pt}
\item[] \hspace*{-\algorithmicindent}\textbf{Input:} The propagation coefficient vector \(\textbf{w}_1\) and matrices \(\textbf{W}_l\), channel matrix \(\textbf{H}^k\), the batch size \(B\), initial learning rate \(\xi _0\), the decay factor \(\gamma\), and the decay interval \(\varpi\).
\item[] \hspace*{-\algorithmicindent}\textbf{Initialization:} The temperature parameter \(T\), the transmission parameters \(\alpha^k_{l,n} \) and \(\theta ^k_{l,n} \) of all SIMs. 
\Repeat 
\State \parbox[t]{\dimexpr\linewidth-\algorithmicindent}{
Performing source encoding for a batch of $B$ input images to obtain \(\mathbf{\Psi}^k_1, k\in\mathcal{K}\).
}
\State \parbox[t]{\dimexpr\linewidth-\algorithmicindent}{
Conducting the forward propagation through (\ref{x}) and computing the classification signal for each sample in the batch by (\ref{s1}) or (\ref{s2}).
}
\State \parbox[t]{\dimexpr\linewidth-\algorithmicindent}{
Calculating the loss function \(\mathcal{L}_B\) by (\ref{loss}).
}
\State \parbox[t]{\dimexpr\linewidth-\algorithmicindent}{
Updating the parameters \(\alpha^k_{l,n}\), \(\theta^k_{l,n}\) and the temperature parameter \(T\) using the Adam optimization algorithm.
}
\State \parbox[t]{\dimexpr\linewidth-\algorithmicindent}{
Updating the learning rate \(\xi_t\) according to the step-based scheduling strategy.
}
\Until{The average loss achieves convergence.}
\item[] \hspace*{-\algorithmicindent}\textbf{Output:} Optimized temperature parameter $T$, the transmission coefficients $\alpha^k_{l,n}$ and $\theta^k_{l,n}$ of all SIMs.
\end{algorithmic}
\end{algorithm}
\endgroup

\vspace{-0.3cm}
\section{Training Process of EMNN}
In this section, we introduce the loss function and the training algorithm for distributed EMNN.

\vspace{-0.3cm}
\subsection{Loss Function}
The softmax function is used to transform the received signal power into a probability distribution of semantic classes. However, unlike conventional digital networks, analog wave-domain computation lacks digital normalization. Coupled with significant wireless channel attenuation, the resulting probability tends to be uniform, hampering gradient propagation during training. Thus, we introduce a trainable temperature parameter $T$ as an physical amplifier to restore the dynamic range of the unnormalized received signals, enhancing the softmax function as $\mathbf{c} =\frac{e^{\hat{\mathbf{y} }/T }}{ {\textstyle \sum_{m=1}^{M}}e^{\hat{y}_{m}/T} } \in \mathbb{R}^{M \times 1}$. The loss function, which measures the discrepancy between the predicted category \(\textbf{c}\) and the ground truth label \(\tilde{\textbf{c}} \), is defined as
\vspace{-0.1cm}
\begin{equation}
\label{loss}
\mathcal{L}\left ( \alpha ^k_{l, n},\theta ^k_{l, n},T \right ) = -\sum_{m=1}^{M} \tilde{\textbf{c}}_{m}\log_{}{\textbf{c}_{m}} ,
\end{equation} 
where $\tilde{\textbf{c}}$ represents a one-hot vector with only the entry corresponding to the image class being $1$.

\vspace{-0.3cm}
\subsection{Training Process}
The purpose of the TAGO algorithm is to minimize the loss function by optimizing the temperature parameter \(T\) and the SIM coefficients, including \(\alpha ^{k}_{l,n}\) and \(\theta ^{k}_{l,n}\), $k \in \mathcal{K}$, \( l \in \mathcal{L}\setminus \left \{ 1 \right \} \), $n \in \mathcal{N}$. The specific steps are outlined as follows:

\textbf{Step 1:}
Initialize the temperature parameter $T = \displaystyle\min_{m \in \mathcal{M}} \hat{y}_m$ based on an arbitrary sample, which dynamically anchors the optimization to the initial channel state by utilizing the actual received power. All SIM parameters are initialized as \(\alpha ^{k}_{l,n}\sim \mathcal{U}\left [ 0,1 \right ] \) and \(\theta ^{k}_{l,n}\sim \mathcal{U}\left [ 0,2\pi \right ) \), \( l \in \mathcal{L}\setminus \left \{ 1 \right \} \), $n\in\mathcal{N}$, where \(\mathcal{U}\) represents the uniform distribution. 

\textbf{Step 2:}
A batch of \(B\) samples is extracted from the image dataset, and the source encoding is performed for each image. 

\textbf{Step 3:}
Perform forward propagation to obtain the predicted category for each sample in the batch through (\ref{x}), (\ref{s1}), and (\ref{s2}). Then, compute the batch loss $\mathcal{L}_B$ via (\ref{loss}).

\textbf{Step 4:}
Iteratively update the transmission coefficients of each SIM and the temperature parameter \(T\) using the Adam optimization algorithm \cite{kingma2014adam}.
 
\textbf{Step 5:} 
To improve the stability of the training process and alleviate the risk of overfitting, we employ a step-based learning rate scheduling strategy, which can be described as $\xi _{t}=\xi _{0} \gamma^{\left\lfloor\frac{t}{\varpi}\right\rfloor}$, where \(\xi _{0}\) and \(\xi _{t}\) denote the initial learning rate and the corresponding value at the $t$-th iteration, respectively. \(\gamma\) indicates the decay factor, while \(\varpi\) is the decay interval.

During the training phase, the GRS computes gradients from the received signals and returns the corresponding gradients to each UAV in designated time slots according to a lightweight coordination protocol. Each UAV then updates its local EMNN parameters, inherently achieving wave-domain channel adaptation by optimizing the phase shifts to counteract specific fading paths. By repeatedly executing Steps 2 to 5, the parameter $T$ and $\mathbf{\Psi}^{k}_{l}$ progressively converge. For clarity, we conclude the procedures in Algorithm 1. 

\begin{figure}[t!]
 \centering
 \begin{subfigure}[t]{0.48\columnwidth}   
   \includegraphics[width=\linewidth]{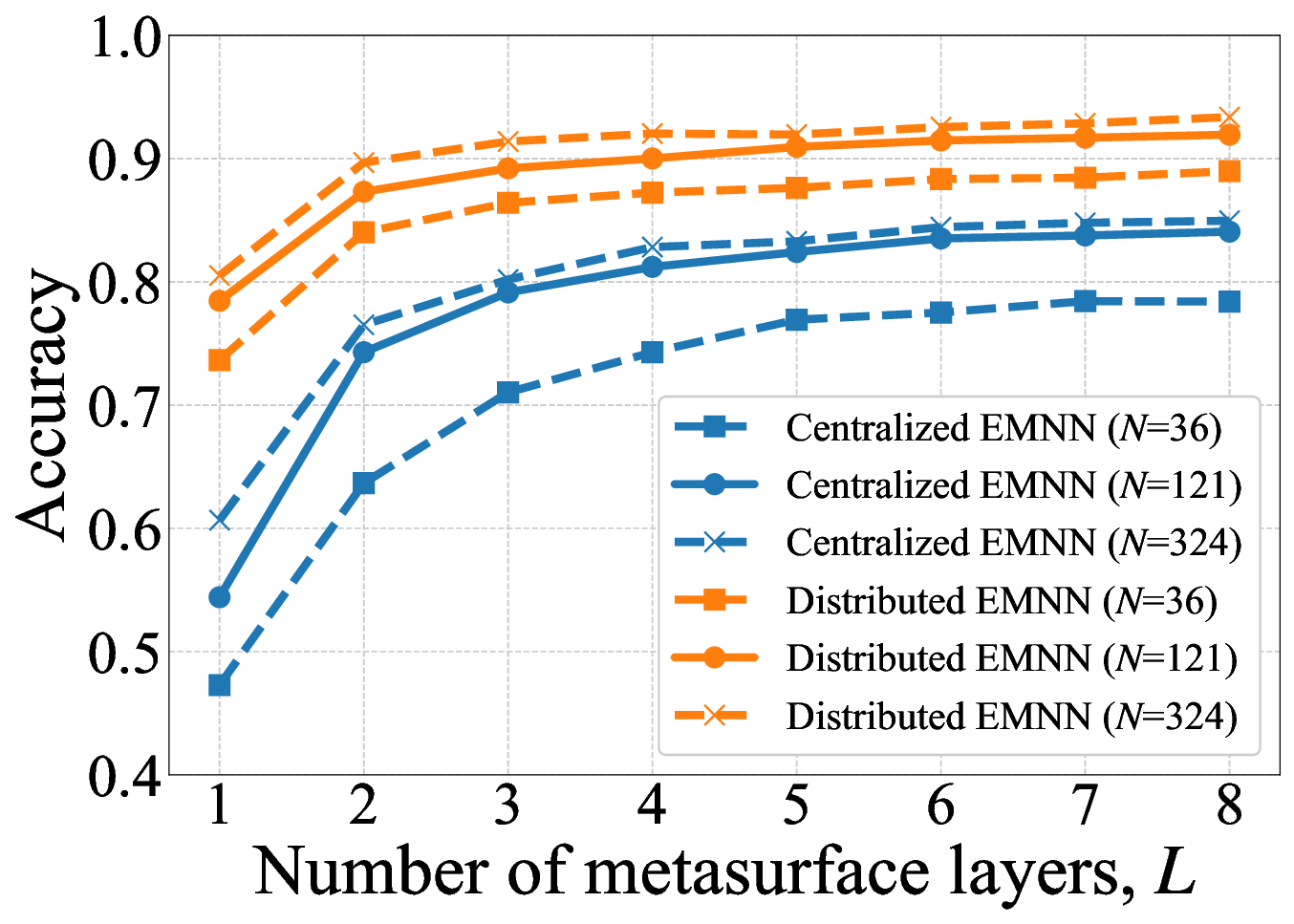}
   \caption{}
   \label{fig:Acc_L}
 \vspace{-0.2cm}
 \end{subfigure}
 \hfill
 \begin{subfigure}[t]{0.48\columnwidth}
   \includegraphics[width=\linewidth]{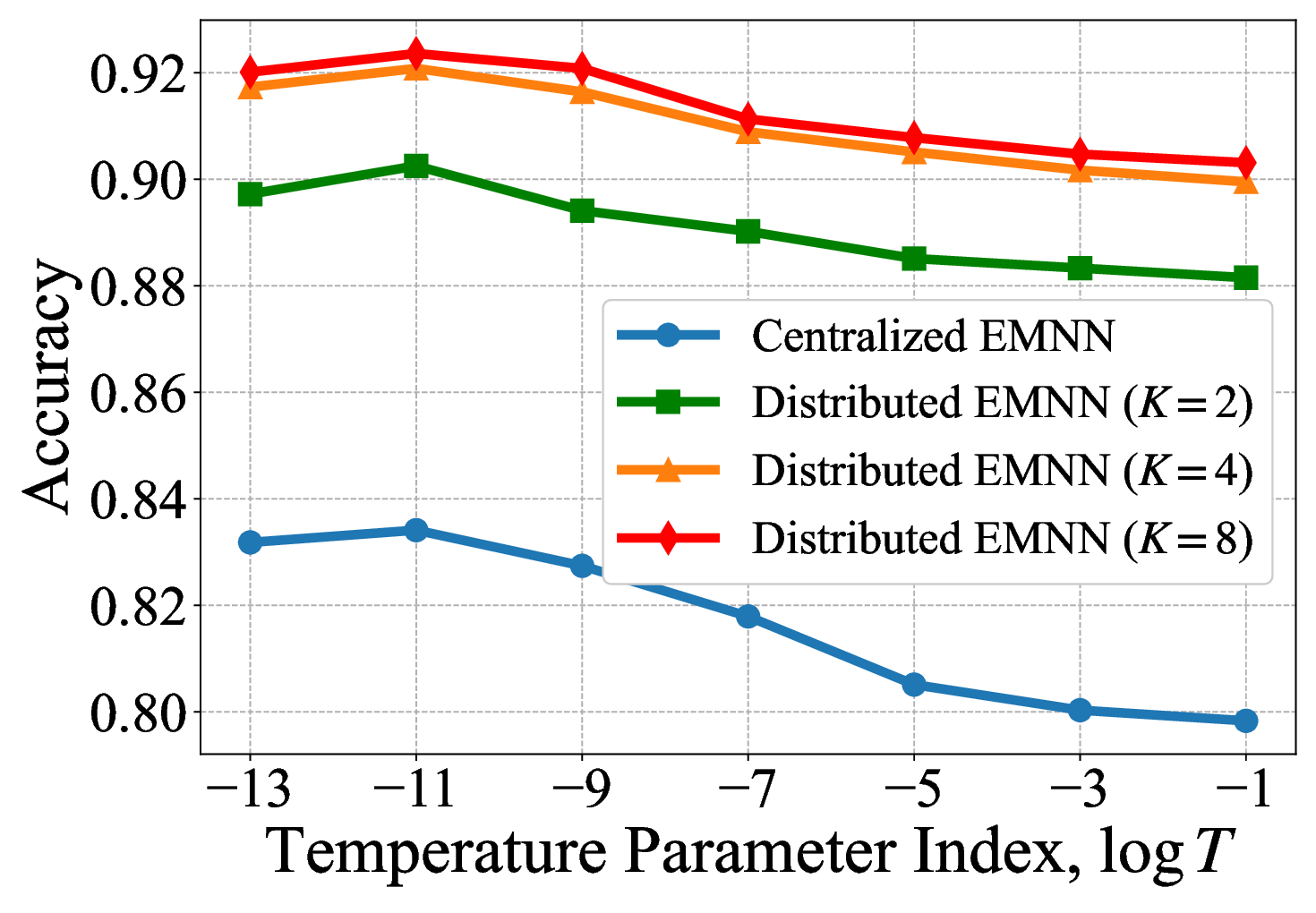}
   \caption{}
   \label{fig:temperature}
 \vspace{-0.2cm}
 \end{subfigure}
 \caption{Performance of EMNN under different system configurations on the MNIST dataset: (a) classification accuracy versus the number of metasurface layers $L$ under centralized and distributed EMNN configurations; (b) classification accuracy versus fixed temperature parameter for EMNN with different numbers of UAVs.}
 \label{fig:performance}
\vspace{-0.6cm}
\end{figure}

\vspace{-0.4cm}
\section{Simulation Results}

\vspace{-0.2cm}
\subsection{Simulation Setup}
Fig.~\ref{fig:system_model} depicts the considered distributed EMNN-based SemCom scenario, where GRS employs \(M=10\) antennas. In the channel, the Rician factor and noise power are set to \(Z_H^k=3\) dB, $k\in\mathcal{K}$ and $\sigma^2=-104$ dBm, respectively. The number of UAVs is set to $K=4$. Each UAV-mounted antenna transmits carrier signals at a frequency of $0.3$ THz, corresponding to a wavelength of \(\lambda =1\) mm, to ensure the SIM and its meta-atoms remain compact enough to meet UAV payload and aerodynamic constraints. The transmit power is set to \(p_t=10\) dBm. All UAVs are randomly positioned around the GRS with respect to azimuth angles \(\mu^k \) and elevation angles \(\beta^k\), and each maintains a fixed propagation distance of \(D_{S, G}=100\) m from the GRS. The path loss \(\Omega^k\) from each SIM to the GRS is set to \(\Omega^k=C_0D_{S,G}^{-\varrho}, k\in\mathcal{K}\), where the pass loss at a reference distance of $1$ m is specified as \(C_0=-35\) dB, with a path loss exponent of \(\varrho=2.5\). Each SIM is composed of \(L=4\) metasurface layers with a thickness of \(T_{\text{SIM}}=10\lambda\). Each metasurface is realized as a UPA consisting of \(N=121\) meta-atoms, uniformly spaced by a distance of \(d_{\text{atom}}=\lambda\), where each meta-atom occupies an area of \(S=\lambda^2\).

In the simulations, the performance of image recognition tasks is evaluated using the MNIST dataset. During training, the input images are rotated at various angles for all evaluated models. Hence, the distributed EMNN concurrently processes differently rotated inputs of the same target to generate distinct entry layers $\Psi_{1}^{k}$ for each SIM, whereas the centralized baseline handles only a single rotated image per inference step. Moreover, an initial learning rate of \(\xi _0=0.01\) and a batch size of \(B=64\) are used, with the decay factor and the decay interval set to \(\gamma=0.8\) and \(\varpi=1\), respectively.

\vspace{-0.5cm}
\subsection{Performance Analysis}
Fig.~\ref{fig:performance}\subref{fig:Acc_L} shows classification accuracy versus the number of metasurface layers $L$ and the number of meta-atoms $N$ per layer, where we consider centralized EMNN with a single SIM and distributed EMNN with multiple SIM configurations. By increasing the number of metasurface layers, the accuracy of all configurations exhibits a tendency to converge. Note that given the same number of meta-atoms, the distributed EMNN consistently outperforms the centralized one by over $10\%$ across different $L$, which is primarily attributed to the distributed EMNN operating at distinct spatial locations collaboratively mitigating the effects of wireless channel fading. Moreover, the classification accuracy of all configurations improves as \(N\) increases, because a larger number of meta-atoms allows for finer-grained image source encoding and greater parameter space, enabling the EMNN to perform more precise wavefront modulation. However, deploying larger SIM configurations inevitably increases the UAV's payload and the static power required for meta-atom control. Thus, designing intelligent energy management strategies to balance wave-domain processing power and system energy limitations is a critical focus for future large-scale EMNN deployments \cite{ameur2025EnergyManagement}.

Fig.~\ref{fig:performance}\subref{fig:temperature} illustrates the classification accuracy of EMNN with different numbers of UAVs under various fixed $T$. As observed, the distributed EMNN consistently outperforms the centralized EMNN, and its performance improves as the number of UAVs increases. Because a larger number of UAV-mounted SIMs provides richer spatial diversity and more reliable aggregation of wave-domain semantic features, enhancing robustness against channel fading. Furthermore, all configurations exhibit a non-monotonic relationship between accuracy and the temperature parameter. This is because when $T$ is too small, the softmax function becomes overly sharp and forces the probability distribution to collapse onto a single category, causing unstable gradient updates and lower accuracy. Conversely, an excessively large $T$ oversmooths the semantic information and flattens the probability distribution, hindering accurate discrimination among categories. Therefore, an appropriate $T$ ensures stable semantic-feature transmission and reliable image recognition performance.

\begin{figure}[t!]
 \centering
 \begin{subfigure}[t]{0.48\columnwidth}
   \includegraphics[width=\linewidth]{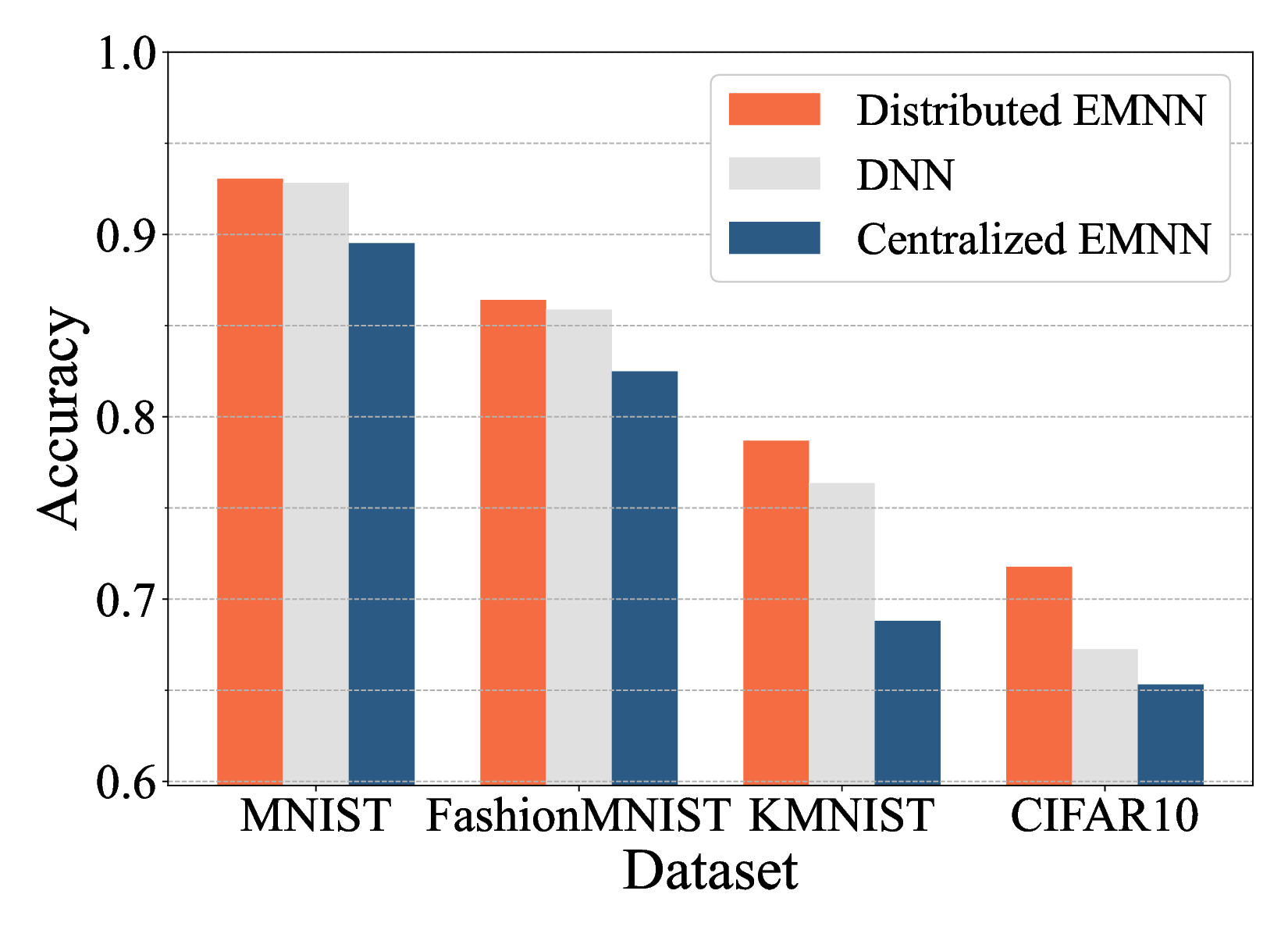}
   \caption{}
   \label{fig:Acc_dataset}
 \vspace{-0.2cm}
 \end{subfigure}
 \hfill
 \begin{subfigure}[t]{0.48\columnwidth}   
   \includegraphics[width=\linewidth]{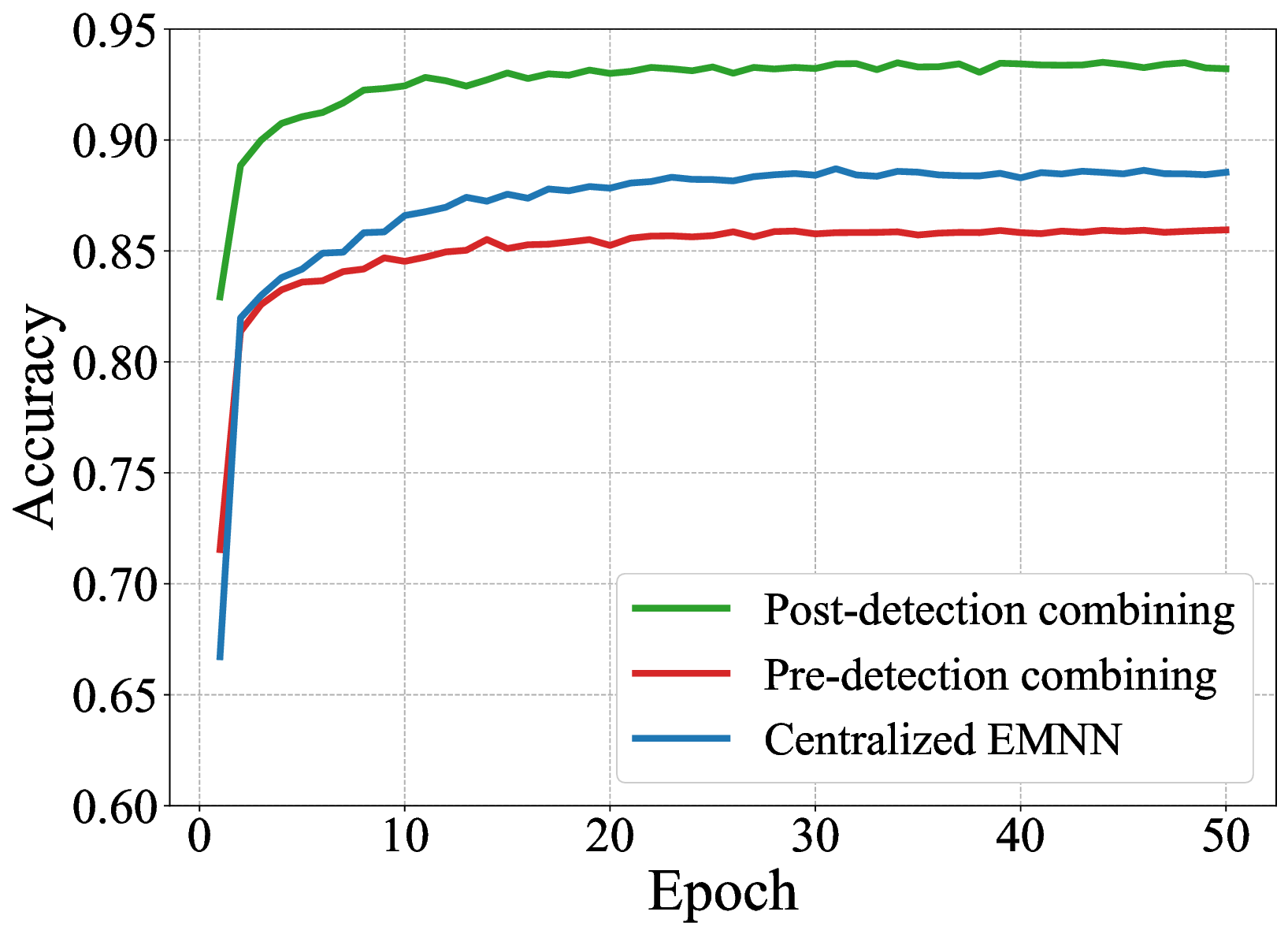}
   \caption{}
   \label{fig:ico_coh_epoch}
 \vspace{-0.2cm}
 \end{subfigure}
 \caption{(a) Classification accuracy comparison on various datasets. (b) Convergence behavior of classification accuracy for different signal combination schemes on the MNIST dataset.}
 \label{fig:convergence and multi-dataset}
\vspace{-0.6cm}
\end{figure}

In Fig.~\ref{fig:convergence and multi-dataset}\subref{fig:Acc_dataset}, we compare the classification accuracy of the proposed distributed EMNN, the centralized EMNN, and a standard deep neural network (DNN) across different datasets. All architectures are constrained to possess an identical total number of trainable parameters, where the centralized EMNN achieves by expanding the number of meta-atoms per layer. Across all datasets, the distributed EMNN consistently achieves the highest accuracy, validating that the performance superiority originates fundamentally from spatial diversity and over-the-air signal fusion rather than parameter expansion. Among these datasets, MNIST yields the highest accuracy of 93.06\%, while CIFAR-10 performs the worst with an accuracy of only 71.77\% for a four-class classification task. This is mainly due to the complexity of the images and the limited number of meta-atoms, which results in a low processing resolution and further degrades the classification performance.

\begin{figure}[t!]
 \centering
 \vspace{-0.3cm}
 \includegraphics[width=0.9\linewidth, trim=20pt 30pt 0pt 90pt, clip]{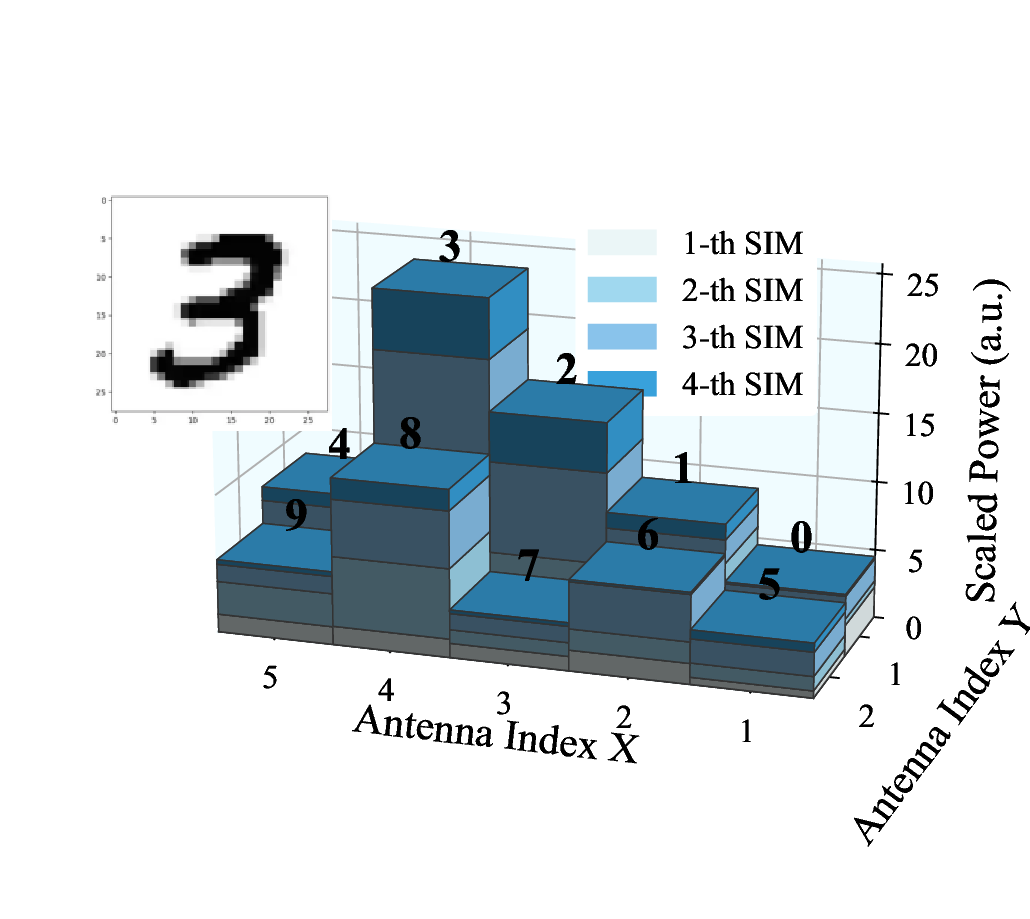} 
 \caption{Scaled power distribution across the receiving antenna array of the distributed EMNN for an MNIST input labeled 3.}
 \label{fig:Power}
 \vspace{-0.6cm}
\end{figure}

Fig.~\ref{fig:convergence and multi-dataset}\subref{fig:ico_coh_epoch} shows the accuracy convergence for parameter-matched architectures. The distributed EMNN with post-detection combining outperforms the centralized one by approximately $5\%$, as over-the-air waveform fusion effectively harnesses spatial diversity to mitigate localized fading. Meanwhile, pre-detection combining underperforms due to signal degradation at the desired antenna from imperfect phase alignment. Moreover, all curves exhibit stable convergence in later training stages, which is attributed to the step-based learning rate decay. Fig. \ref{fig:Power} illustrates the scaled power distribution across the antenna array at the GRS for an input image labeled 3, where scaled power refers to temperature-scaled energy output from the EMNN. The four colors represent the individual power contributions from different SIMs. The antenna indexed as 4 (corresponding to digit 3) exhibits the highest scaled power, indicating that the distributed EMNN focuses the learned energy toward the target antenna. Moreover, distributed deployment mitigates channel fading by allowing UAVs with better channels to transmit semantic information, enabling adaptive aggregation and enhancing reliability.

\vspace{-0.5cm}
\section{Conclusion}
In this letter, we proposed a task-oriented SemCom system based on UAV-enabled distributed EMNNs. Unlike digital neural networks, EMNNs directly perform semantic processing in the wave domain to substantially reduce latency and energy consumption. Moreover, the distributed design collaboratively encodes semantic information to improve robustness against wireless channel fading. To support training, the introduced TAGO algorithm adaptively tunes a learnable temperature parameter to enhance recognition performance. Simulation results validate that the distributed EMNN consistently outperforms the centralized baseline across multiple datasets. Future research may explore integrating energy harvesting for power-limited UAVs, jointly optimizing UAV and SIM power for ultra-low-power EMNNs, and developing adaptive strategies to mitigate mobility-induced phase shifts during AirComp.

\vspace{-0.4cm}
\bibliographystyle{IEEEtran} 
\bibliography{An_WCL2026-1139}

\newpage
\vfill
\end{document}